\documentclass[aps,prl,groupedaddress,twocolumn]{revtex4-2}

\usepackage{color}
\usepackage{lineno}
\usepackage{graphicx}

\begin{document}


\title{On the Origins of Tension–Compression Asymmetry in Crystals \\and Implications for Cyclic Behavior}


\author{Sylvain Queyreau}
\affiliation{Universite Sorbonne Paris Nord, LSPM-CNRS, UPR 3407, 93430 Villetaneuse, France}
\author{Benoit Devincre}
\affiliation{Université Paris-Saclay, LEM, CNRS - ONERA, F-92322 Chatillon, France}


\date{December 25$^{th}$ 2020}

\begin{abstract}
Most of crystalline materials exhibit a hysteresis on their deformation curve when mechanically loaded in alternating directions. This Bauschinger effect is the signature of mechanisms existing at the atomic scale and controlling the materials damage and ultimately their failure. Here, three-dimensional simulations of dislocation dynamics and statistical analyses of the microstructure evolution reveal two original elementary mechanisms. An asymmetry in the dislocation network junctions arising from the stress driven curvatures and the partial reversibility of plastic avalanches give an explanation to the traction-compression asymmetry observed in FCC single-crystals. These mechanisms are then connected in a physically justified way to larger-scale representations using a dislocation density based theory. Parameter-free predictions of the Bauschinger effect and strain hardening during cyclic deformation in different materials and over a range of loading directions and different plastic strain amplitudes are found to be in excellent agreement with experiments. This work brings invaluable mechanistic insights for the interpretation of experiments and for the design of structural components to consolidate their service life under cyclic load.
\end{abstract}


\maketitle


The materials mechanical response depends upon the deformation history. At the macroscale, this manifests itself in the shape of hysteresis loops on force-displacement curves when the loading direction is alternated \cite{Miguel:2006ly}. Thus, an initial deformation in tension usually facilitates a reverse deformation in compression, at least over a certain transient. This phenomenon, known as the Bauschinger Effect (BE) is intimately related to fundamental aspects of material behavior such as strain hardening and life prediction under cyclic loading. The issue of the durability of crystalline materials in non-monotonic solicitation is a key problem in a large variety of applications ranging from microelectronic devices to biocompatible implants and monumental structural frames of buildings \cite{Kubin:2013fk}. It is consequently associated with crucial technological, economical and safety stakes. Modelling the BE and cyclic behavior is a long-standing problem and a physically based model with real predictive capability has yet to be formulated. Here, we elucidate the puzzling and most argued BE observed in FCC single crystals and quantify its implication for cyclic behavior. This step is essential before addressing more  complex polycrystalline materials.

Dislocation dynamics and their interactions are key phenomena controlling the plasticity of the crystalline materials and the formation of organised microstructures. At the small scale, their intermittent and mostly irreversible motion presents similarities with avalanche phenomena \cite{Dimiduk:2006, Csikor:2007lr, Devincre:2008lr}. Until now, tension-compression asymmetry was thought to find its origin in the building up of a resistive long-range backstress during the first loading \cite{Asaro:1975eu, Mughrabi:88} or easy slip after reversal of the strain induced by partial dissolution of the dislocation network \cite{Buckley:1956bh, Sleeswyk:1978cr, Depres:2008fk, Rauch:2011kx}. In principle, these mechanisms can facilitate dislocation motion during reverse strain, but neither any known elementary dislocation mechanism nor any in-situ experimental evidence support them in single crystals. Moreover, they do not fit with the embraced 'forest' model \cite{Devincre:2008lr, Fried:67} according to which the formation of dislocation junctions controls plastic flow and strain hardening.

\begin{figure*}[htbp]
\includegraphics[width = 0.25\linewidth]{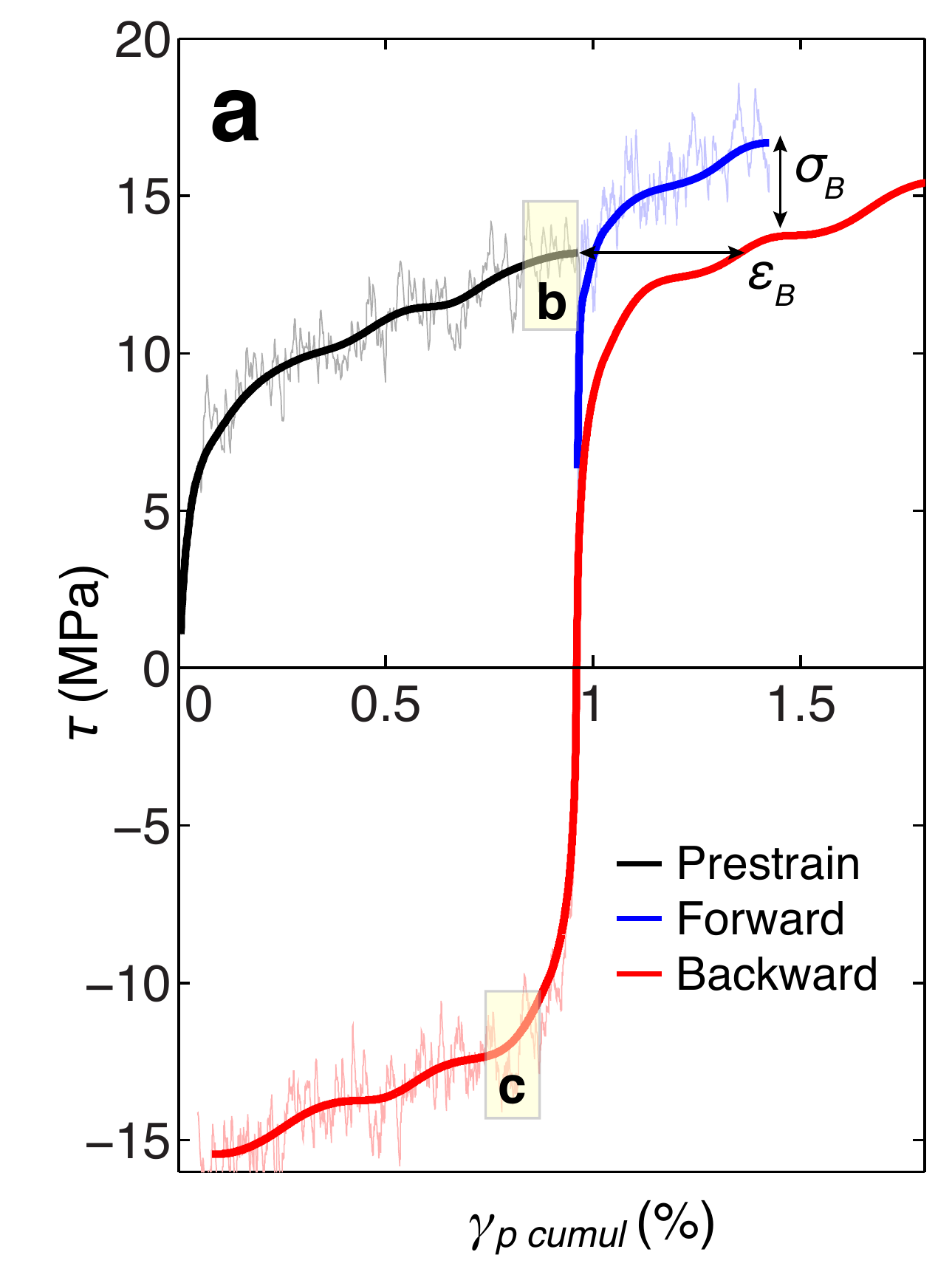}
\includegraphics[width = 0.73\linewidth]{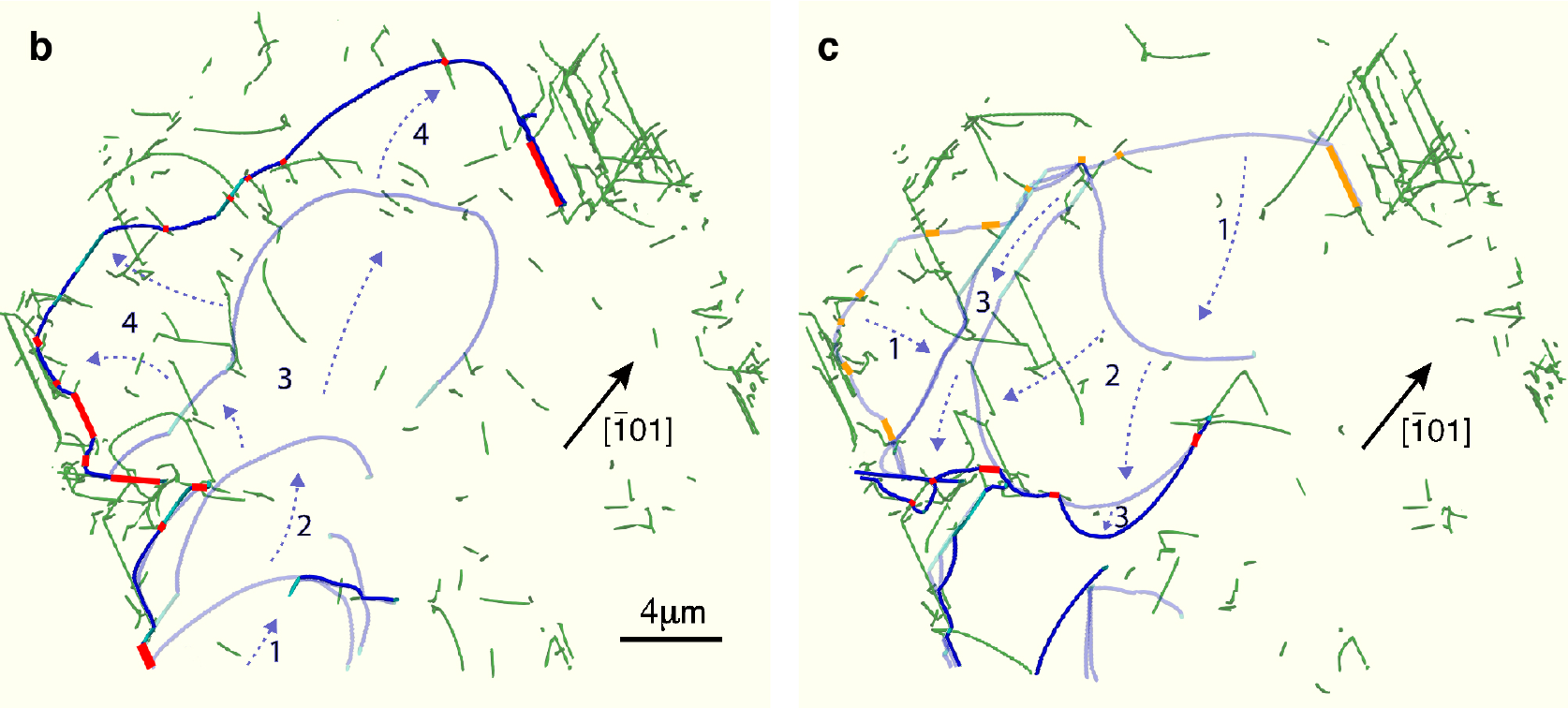}
\caption{Typical hysteresis curve and dislocation microstructure during a DD simulation of a Bauschinger experiment of a Ni single crystal oriented along $[001]$ axis.  a, Hysteresis curve representing the flow stress as a function of cumulated plastic strain. Prestrain in tension (in black) is followed by either a compression (in red) or continued tension (in blue). Thick lines describe the average behavior calculated from a Gauss kernel smoothing procedure. The Bauschinger stress $\sigma_B$ and strain $\epsilon_B$ are defined on this figure. b, The unzipping of a junction triggers a plastic avalanche at the end of the prestrain ($0.8-1\%$). This sequence is constructed from periodic snapshots of the dislocation network in a slice of material ($0.1\;\mu \mathrm{m}$ thick) where a dislocation avalanche takes place. A mobile dislocation (in blue) forms junctions (thick red lines) with ‘forest’ dislocations (in green) cutting the slice. c, The same slice observed at the beginning of compression ($1 - 0.8\%$). Segments of dislocation break away from weak junctions and progressively remobilize a long section of dislocation leading to the loss of many other junctions (in orange). The mobile dislocation sweeps areas that were first visited during tension and set free many ‘forest’ dislocations anchored by junctions formed during the prestrain.}
\label{fig:1}
\end{figure*}

To shed light on this problem, tension-compression tests were performed with 3D Dislocations Dynamics (DD) simulations. We find that the hysteresis behavior of plastic deformation originates from two dislocation mechanisms observed after the loading reversal. From averages performed on the collective behavior of dislocations, we define a crystal plasticity model that captures, without any fitting parameter, most of the features of the BE and of the cyclic strain hardening observed in FCC single crystals. 

All simulations are performed using the DD simulation code microMegas \cite{Devincre:2011fk}. Periodic boundary conditions are applied to ensure that the dislocation avalanches occur without amplitude restriction, in agreement with the plastic processes found in single crystals or large grains in polycrystals. Cu and Ni are taken as reference materials and for comparison with experimental data. Having a systematic approach, our simulations cover an extensive set of experimental configurations where key parameters are varied, namely, the initial dislocation density ($10^{11}-10^{15}\,\mathrm{m}^{-2}$), the amount of plastic strain accumulated during prestrain ($0.25-8\%$) and crystal orientation ($[135]$, $[112]$, $[111]$ and $[001]$). Detailed information on the methods, including statements of data availability and any associated accession codes and references, are available in the joint publication \cite{Queyreau:2021}.

First, we show that DD simulations reproduce the main features of tension-compression asymmetry observed experimentally. Choosing the [001] deformation axis as a representative example of other crystal orientation, Fig.~\ref{fig:1}a shows a typical stress-strain curve obtained during a Bauschinger cycle simulation. Several slip systems, associated with different slip planes and different Burgers vectors, are simultaneously shearing the crystal. Their interactions induce a dislocation storage and an increase in the flow stress with a linear slope close to $\mu/150$ (where $\mu$ is the material shear modulus) in good agreement with experiments \cite{Kubin:2013fk}. Serrations are clearly visible on the simulated deformation curves and can be attributed to strain bursts and dislocation avalanches. Serrations are also present on the compression curves and continued tension second loadings. The reverse compression curve exhibits a reduced elastic limit, a well-rounded appearance of the initial plastic portion (characterized by a strain shift $\epsilon_{BE} \approx 0.4\%$) and a permanent stress softening ($\sigma_{BE} \approx 3\,\mathrm{MPa}$) with respect to the continued tension curve.

While the many BE tests we made compare well with experiments, the simulation analysis surprisingly does not support usual interpretations of the BE. For instance, no significant backstress is found in the simulated volumes at the beginning of the reversed deformation when the applied loading is still close to zero. This finding agrees with recent Xray microdiffraction measurements \cite{Kassner:2009fk, Kassner:2013fk}, where the residual stress induced by a dislocation microstructure formed in a tensile test was found as low as a few percent of the flow stress.

Actually, our simulations give evidence that the BE is caused by two original mechanisms. As illustrated in Fig.~\ref{fig:1}b, plastic deformation proceeds through intermittent dislocation motion sequences. Each sequence starts with the destruction of a junction with the lateral \emph{unzipping} of a mobile dislocation progressing under the effect of external loading \cite{Bulatov:98}. This first event triggers a cascade of other junction destructions and significant dislocation displacements characterized by a scale-free power law distribution \cite{Devincre:2008lr}. During such bursts in dislocation dynamics, new junctions can be formed each time a mobile dislocation collides with a ‘forest’ dislocation (an immobile dislocation piercing the mobile dislocation glide plane) if this reaction minimises the elastic energy. As illustrated in Fig.~\ref{fig:1}b, these plastic strain avalanches leave in their wake a large amount of immobile dislocation segments at the origin of dislocation density increase and material strengthening. Each avalanche stops when the whole mobile dislocation length is pinned on the dislocation network.

The same zone observed during the later compression (Fig. ~\ref{fig:1}c) shows a very different behavior. First, former stable junctions that anchored the dislocation network at the end of the tensile loading are now easily unzipped. A first tension-compression asymmetry is therefore found in the junctions stability. This effect is unexpected because it contrasts with basic calculations that consider only two interacting dislocations. Here, the junction weakening we see in the simulations comes from the interactions between several forest dislocations and a curved mobile dislocation. Indeed, arrangements of three consecutive junctions zipped along a curved dislocation are less (respectively more) stable than isolate junctions formed by straight segments as an effect of an additional line curvature energy working to decrease (respectively expand) the junction length. 

\begin{figure}[htbp]
\includegraphics[width = 0.75\linewidth]{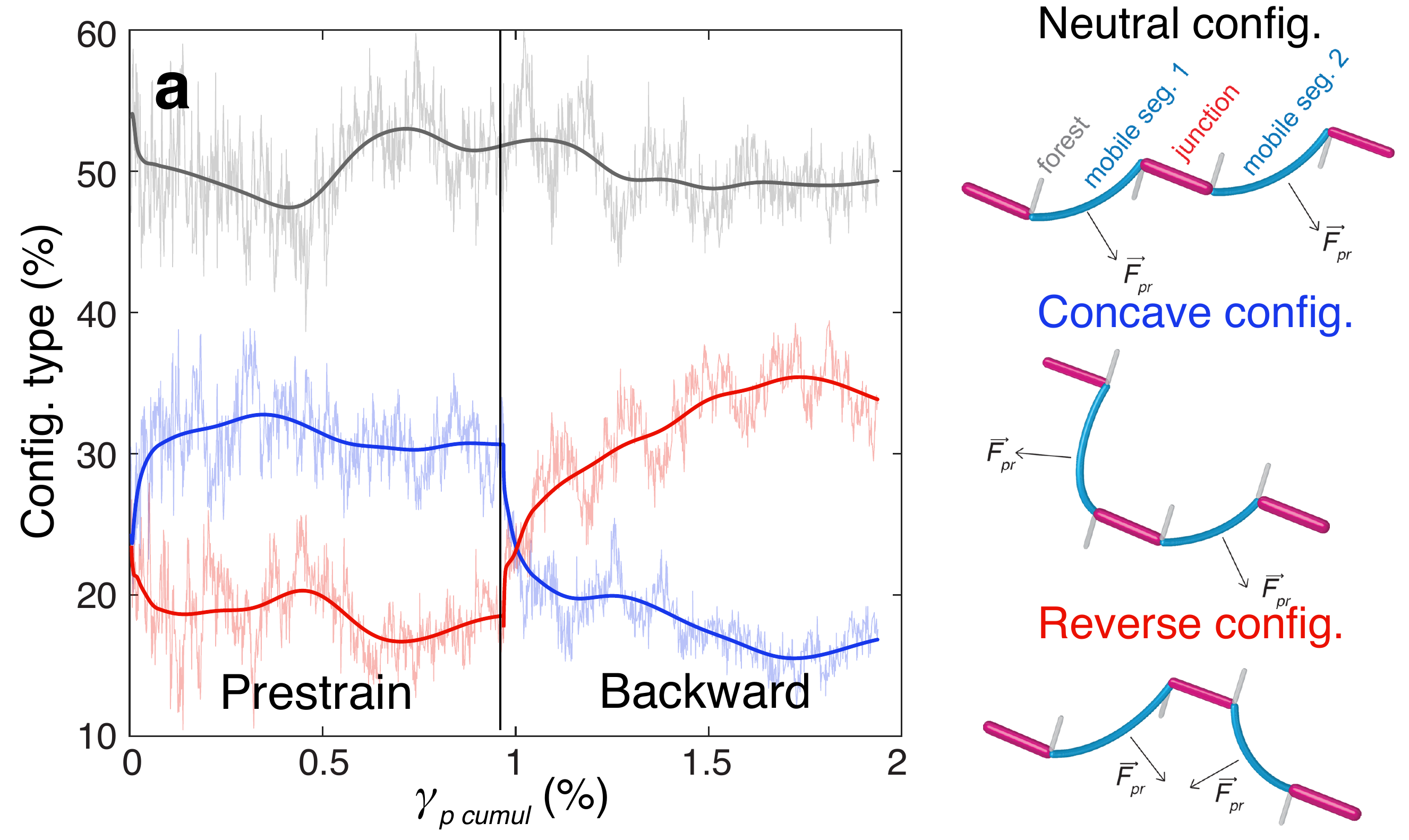} \\
\includegraphics[width = 0.6\linewidth]{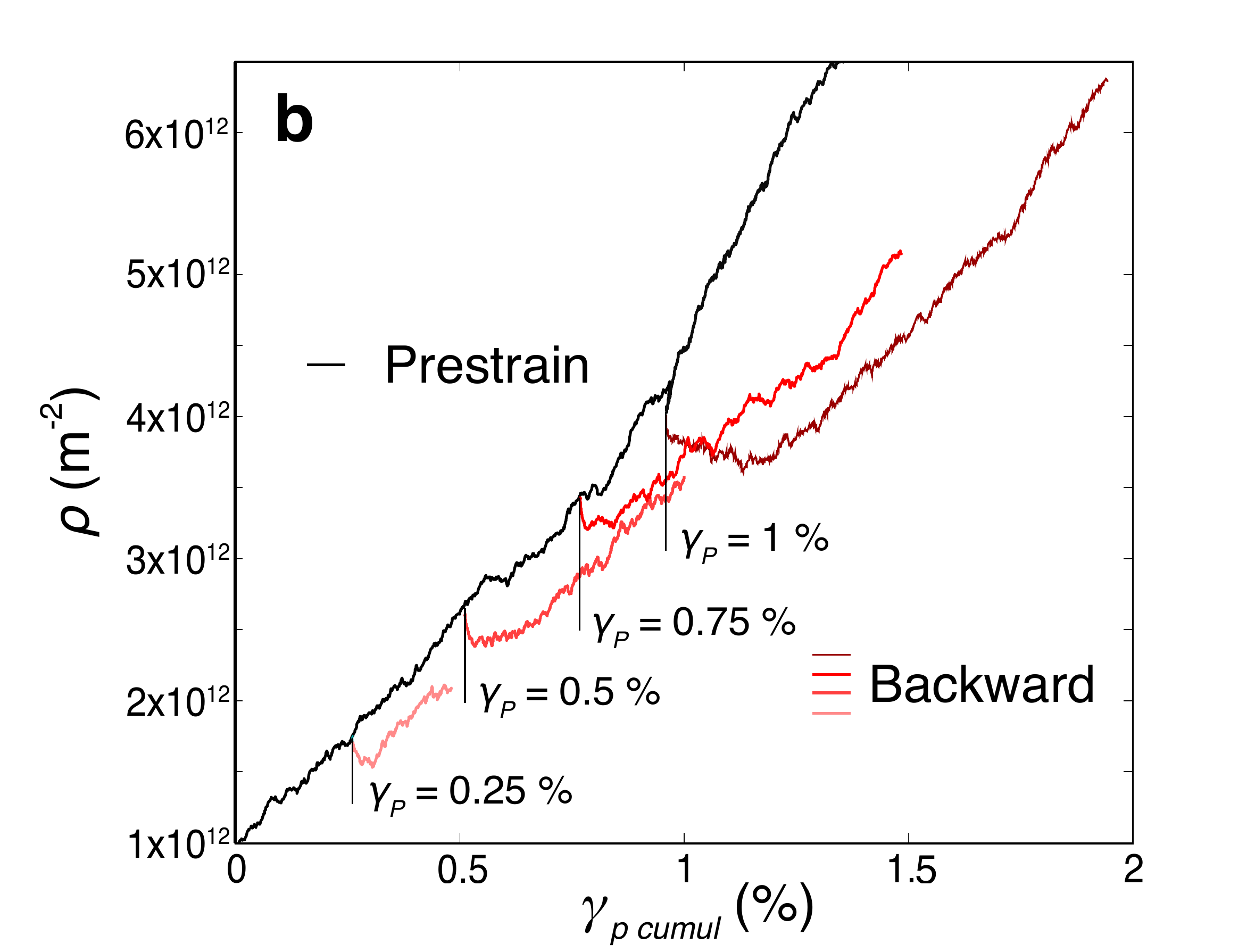}
\caption{Evidence of the two original mechanisms at the origin of the traction-compression asymmetry. a, Statistical analysis of the many configurations of junctions found in the 3D dislocation network and formed during the $[001]$ Bauschinger test simulation. The microstructure is analyzed in terms of ternary configurations with different mechanical stability depending on the loading direction. The concave, reverse and neutral configurations are defined in the text. The simulated microstructure contains more than two thousand junctions. b, Drop in the dislocation density observed at loading reversal as a function of the prestrain amplitude. The black curve displays the ‘regular’ dislocation storage associated with continued tension.}
\label{fig:2}
\end{figure}

When considering the large variety of dislocation arrangements existing in simulated networks, one can distinguish three configuration geometries (see schematics on Fig.~\ref{fig:2}a), (i) concave configurations where two segments are pulling a given junction backward due to the spatial arrangement of other dislocations, (ii) reverse configurations where two segments are both pulling the connected junction forward and (iii) neutral configurations where two segments are pulling the considered junction in opposite directions. Concave configurations (i) are, on average, more mechanically stable than reverse configurations since mobile segments ending at the junction promote the zipping rather than the unzipping of the junction. This is why, as illustrated in Fig.~\ref{fig:2}a, concave configurations represent a larger number ($\approx 32\%$) of the configurations found in the dislocation networks than reverse configurations ($\approx 18\%$), while neutral configurations take the rest ($\approx 50\%$). When loading is changed from tension to compression, concave (i) and reverse configurations (ii) switch roles as the direction of dislocation curvature is reversed. This explains the excess of weaker reverse configurations that is systematically observed at the beginning of backward deformation. This abnormal partition of junction configurations progressively disappears as deformation progresses.

A second and stronger asymmetry in the tension-compression behavior is found in the kinetics of dislocation glide. The backward motion sequence of Fig.~\ref{fig:1}c shows that deformation still occurs through slip bursts. However, plastic events now take place within regions that have already been explored by mobile dislocations during their recent forward motion. In turn, avalanches at the beginning of the backward deformation eliminate many junctions previously formed in the dislocation network. This effect is somewhat similar to what is observed in alloys \cite{Queyreau:2009lr}, where mobile dislocations collide with dislocation loops formed during prestrain. Consequently, as illustrated in Fig.~\ref{fig:2}b, a significant decrease of the dislocation density is noticed at strain reversal. Due to the irreversible nature of plastic events, dislocations will eventually explore new material regions and this effect will wear off with deformation.

With the ambition in mind to connect with larger scale simulations, we now formulate constitutive equations that describe the mechanisms discovered in the DD simulations. Since the plastic behaviour observed in our BE test simulations is still controlled by forest interactions and plastic bursts leading to dislocation storage, we consider a dislocation density based theory of Crystal Plasticity (CP) established by the seminal work of Kocks and Mecking \cite{Kocks:03, Teo:93, Kubin:2008fk}. Next, we expose the main modifications required to account for the BE within this framework. The model is presented in its entirety in the joint publication \cite{Queyreau:2021}.

\begin{figure}[htbp]
\includegraphics[width = 0.6\linewidth]{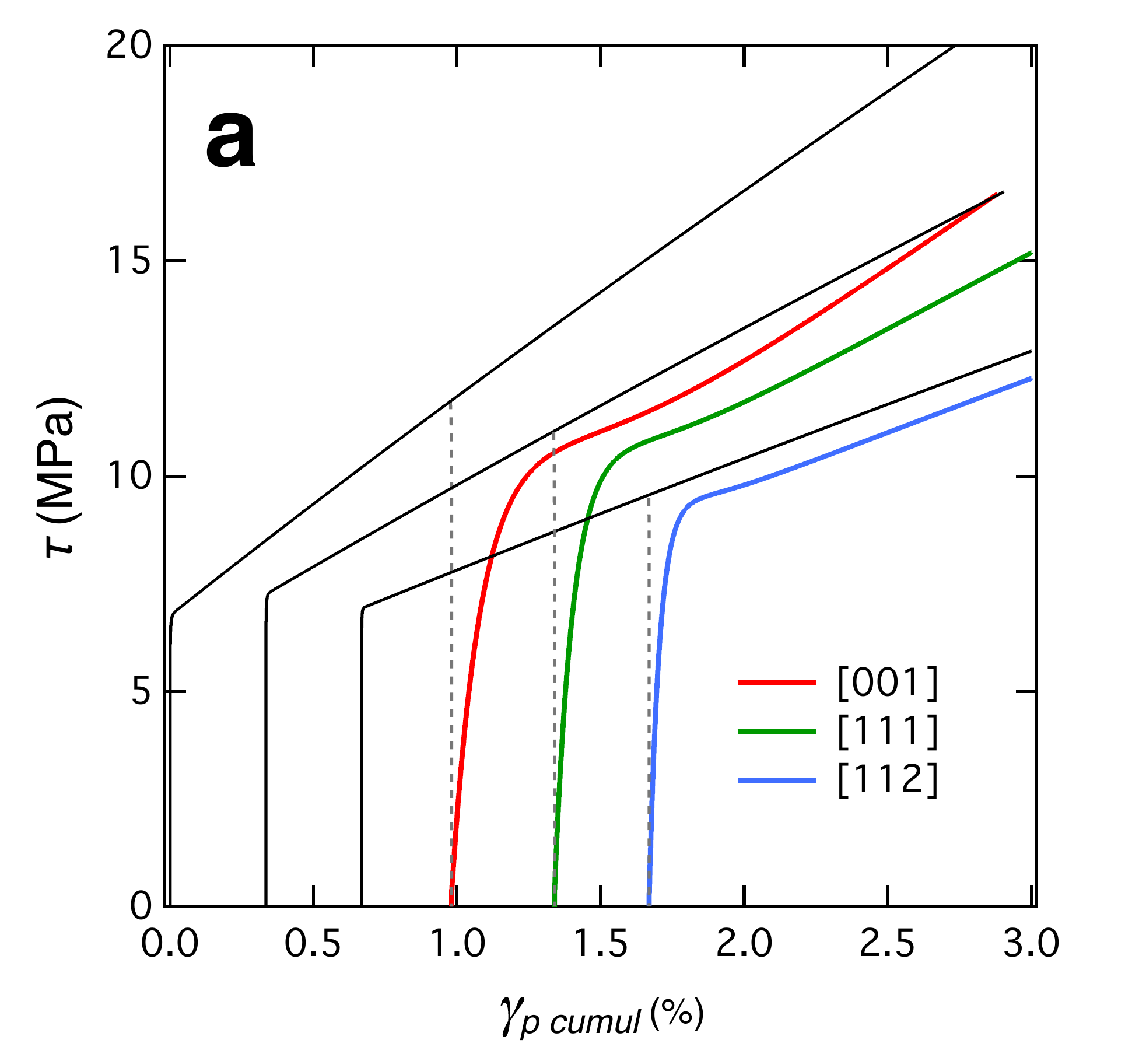} \\
\includegraphics[width = 0.6\linewidth]{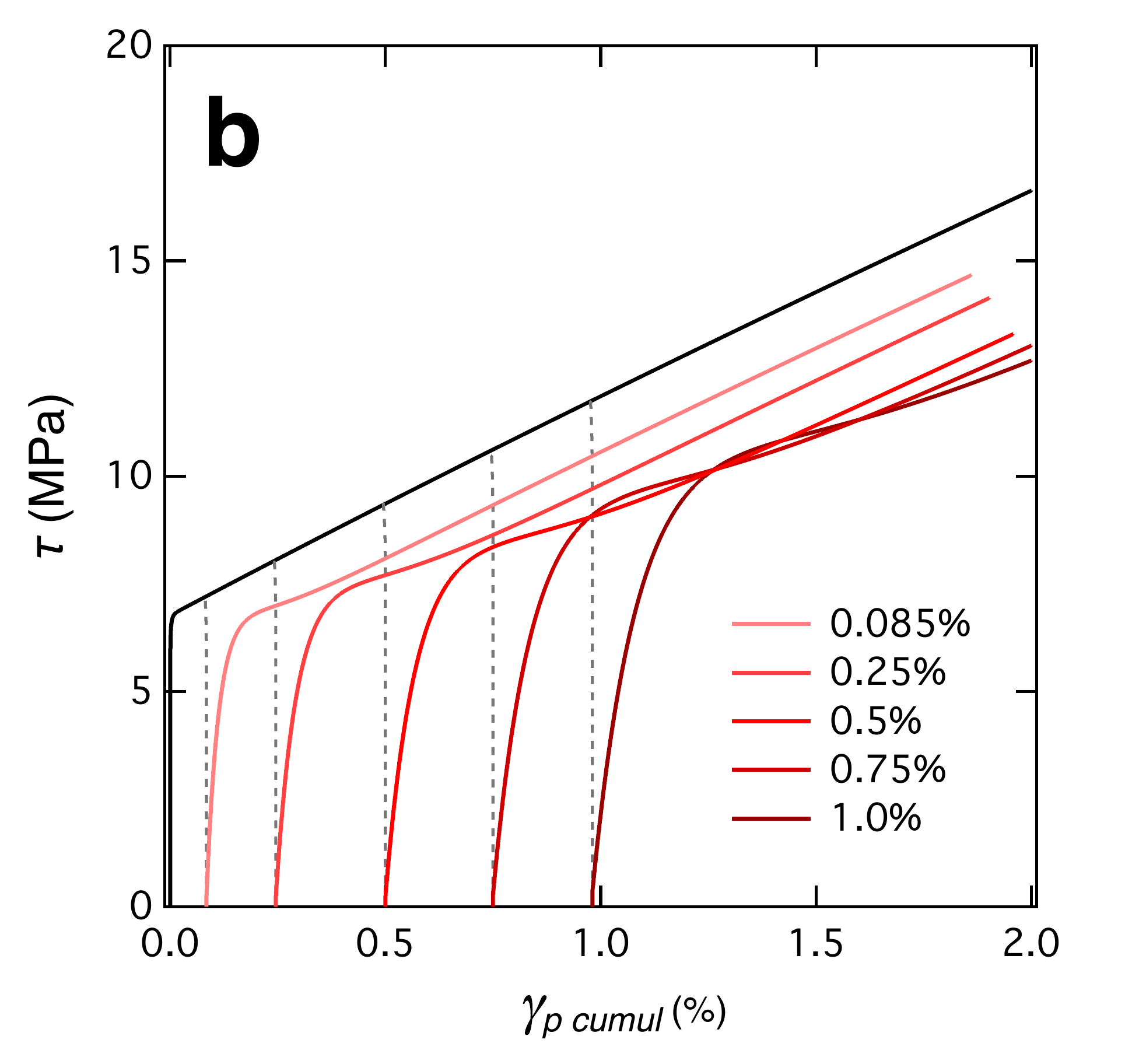}
\caption{Simulations of BE tests with our CP model physically justified from the DD simulation results. Continued tension curves in black are compared with subsequent compression curves in color. a, Influence of the loading orientation on the hysteresis curves for a Ni single crystal deformed along high symmetry axis after identical prestrain. b, Effect of an increasing prestrain on the following compression curves for a $[001]$ deformation.}
\label{fig:3}
\end{figure}

Junction stability appears in the model within the flow stress equation and is accounted for with an interaction matrix whose coefficients $a_{ij}$ measure the average junction strength between an active slip system ‘i’ and a forest slip system ‘j’. $a_{ij}$ dimensionless coefficients depend mostly upon the type of reactions and are well known from experiments or simulations in the case of monotonous loadings. To reproduce the weakening of junction stability induced by the deformation history, we write:

\begin{equation}
a_{i j}^{b c k} = ( 1 - r_a ) \times a_{i j} \mathrm{,~~with~~}\;r_a = \exp ( \frac{- \gamma_{b c k}^i} {C_a b \sqrt{ \Delta \rho_{pr}^i}} )
\label{eq:newa}
\end{equation}

\noindent where the ‘pr’ and ‘bck’ subscripts refer to prestrain and backward deformation, respectively. The term within the exponential states the competition between the easy destruction of existing junctions formed during previous deformation stage and the formation of fresh junctions in newly explored areas of the crystal. $\Delta \rho_{pr}^i$ is the amount of density stored during the first loading and is proportional to the number of junctions formed \cite{Devincre:2008lr}. $C_a$ is a dimensionless coefficient describing the length of the transient regime.

The Mean Free Path (MFP) of dislocations, which describes the distance covered by dislocations before their temporary or permanent immobilisation, controls the storage rate of dislocation density. Dislocation MFP depends upon few physical parameters among which is the rate of formation of immobilizing junctions per unit of strain, $p_0$ \cite{Devincre:2008lr}. Hence, the loss of stored dislocations corresponds to a lengthening of the dislocation MFP. The evolution of $p_o$ revealed by DD simulations after stress reversal is accounted for by the following equation:

\begin{equation}
p_{0}^{b c k} = ( 1 - r_p ) \times p_{0} \mathrm{, with}\; r_p = A_p \times \exp ( \frac{- \gamma_{b c k}^i} {C_p b \sqrt{ \Delta \rho_{pr}^i}} )
\label{eq:P0bck}
\end{equation}

\noindent where $A_p$ and $C_p$ are two dimensionless constants that measure the initial drop and the length of the transient regime on $p_0$, respectively. The three coefficients $C_a$, $A_p$ and $C_p$ were unambiguously evaluated from a large set of different DD simulations leading to $C_a = 0.6 \pm 0.1$, $A_p = 2.2 \pm 0.6$ and $C_p = 2.5 \pm0.3$ see the joint publication \cite{Queyreau:2021} for calculation details).

\begin{figure}[htbp]
\includegraphics[width = 0.75\linewidth]{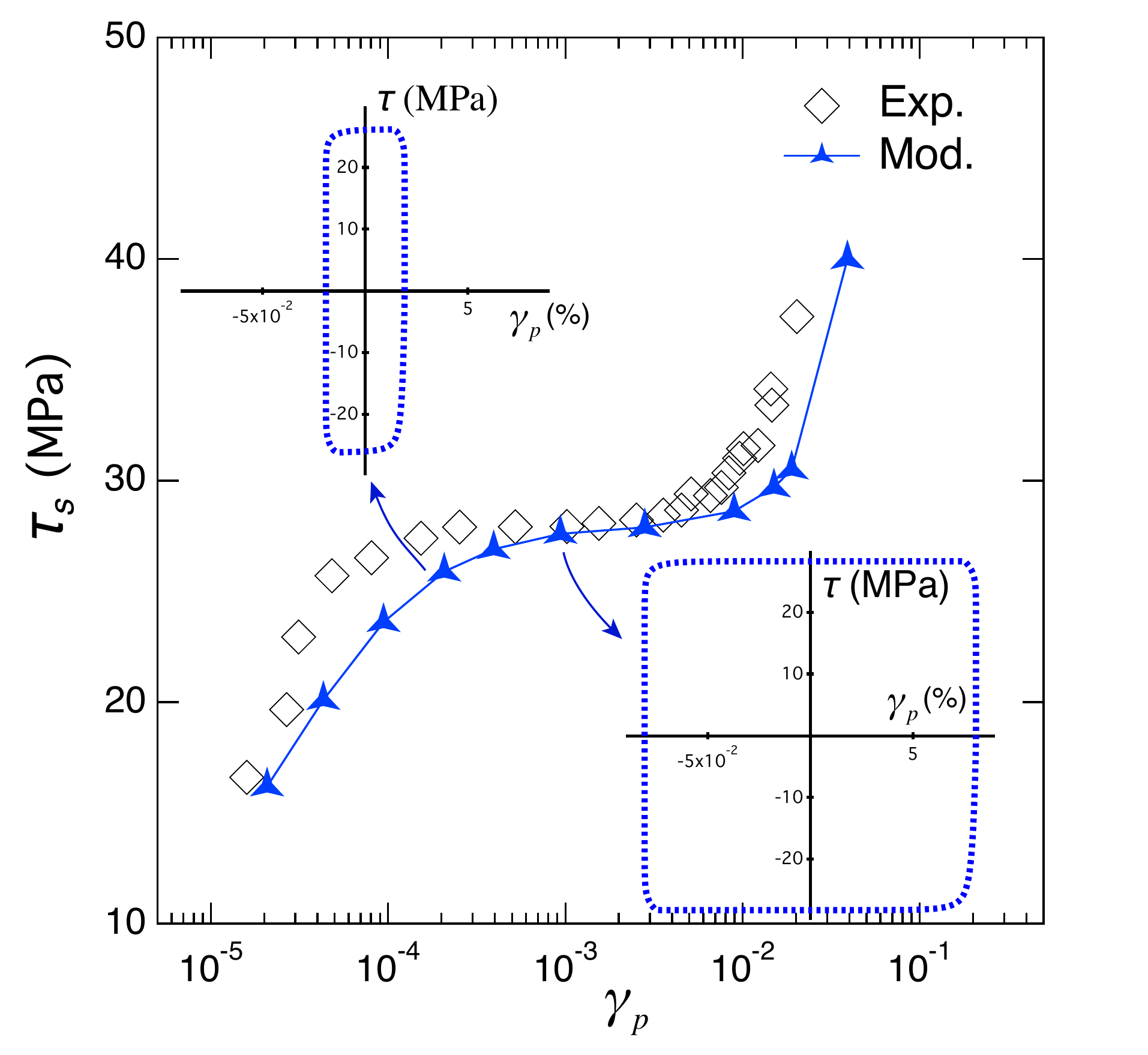}
\caption{Illustration of the prediction capability of our CP model when applied to cyclic deformations (in blue). As the number of cycle increases, hysteresis curves loose their roundness, until reaching saturation when no hardening is observed. The saturation stress $\tau_s$ exhibits a complicated dependence upon imposed strain per cycle with the existence of three stages perfectly predicted by the model. The hysteresis curve at saturation is also well reproduced (see insets). Experimental data (squares) correspond to the noteworthy work from \cite{Mughrabi:1978uq} on Cu single crystals deformed in single glide conditions.}
\label{fig:4}
\end{figure}

To compare with existing experimental data on single crystals, we solved the equations of the CP model including equations \ref{eq:newa}-\ref{eq:P0bck} with the Finite Element Method (FEM) where the relevant physical fields (see supplementary materials) are now continuous and defined at the scale of macroscopic specimens. A selection of obtained deformation curves is given in Fig.~\ref{fig:3}. These results agree well with DD simulations and experiments. Those CP calculations provide additional useful insights on the BE. The observed increase of the BE stress and BE strain with the amount of prestrain \cite{Buckley:1956bh, Pedersen1981} is the outcome of the increase of $\Delta \rho_{pr}$, the dislocation density stored during prestrain. Also, the BE intensification observed with the increasing number of active slip systems \cite{Buckley:1956bh, Yakou:1977aa, Wadsworth:1963aa, Marukawa:1971aa} is explained by a larger fraction of dislocation density that is affected by equations \ref{eq:newa}-\ref{eq:P0bck}. Last, single glide orientation $[135]$ also exhibits a BE, but is relatively smaller in intensity \cite{Buckley:1956bh, Yakou:1977aa, Wadsworth:1963aa} and is due to the very low rate of dislocation storage for this orientation.

Without the use of any ad hoc parameters or backstress definition, our CP model also captures most of the features of the cyclic behavior of single crystals submitted to alternating tension and compression. This is shown in Fig.~\ref{fig:4}, where the cyclic behavior of Cu single crystals deformed in [135] single glide conditions is compared with experimental data \cite{Mughrabi:1978uq}. With the accumulation of plastic strain, flow stress $\tau_c$ increases at every cycle very quickly at first, but ultimately saturates for large amounts of cumulated plastic strain. This saturation stress $\tau_s$ is controlled partly by the new dislocation processes identified for the BE and partly by the more conventional dynamic recovery mechanisms already present in monotonic deformation \cite{Kubin:2008fk}. The model captures quantitatively the complex evolution of the saturation stress with the increase of plastic strain amplitude per cycle, and the loss of roundness of hysteresis curves at saturation. Saturation stress $\tau_s$ first increases with the strain $\gamma_p$ up to $10^{-4}$, and then exhibits a plateau behavior for larger strain increments between $10^{-4}$ and $10^{-2}$ where materials response is dominated by recovery processes (see joint publication \cite{Queyreau:2021} for more details). A third stage is ultimately triggered by the activation of a secondary slip system.

To summarize, thanks to a multiscale modeling approach we identified the key mechanisms controlling the tension-compression asymmetry in FCC single crystals. From this, a crystal plasticity model free of adjustable parameters was proposed. This model reproduces the BE and the cyclic deformation of FCC single crystals in a physically justified manner. These results pave the way for new predictive models of cyclic deformation, and ultimately for the understanding of complex behavior like plastic strain localisation in devices which are key issues to assess the service life of materials.

\bibliography{Bauschinger.bib}

\end{document}